\documentclass[twocolumn]{aa}
\usepackage{graphicx}
\usepackage{amssymb}
\usepackage{txfonts}
\begin{document}

\title{Modeling of the heliospheric interface: multi-component nature of the heliospheric plasma}
\author{Yury G. Malama\inst{1,3}, Vlad V. Izmodenov \inst{2,3}, Sergey V. Chalov \inst{1}}

\authorrunning{Malama et al.}
\titlerunning{Multi-component nature of the heliospheric interface}
\offprints{V.V. Izmodenov} \institute{Institute for Problems in
Mechanics, Russian Academy of Sciences \and Lomonosov Moscow State
University, Department of Aeromechanics, School of Mechanics and
Mathematics \& Institute of Mechanics, Moscow, 119899, Russia
\email{izmod@ipmnet.ru}
 \and Space Research Institute (IKI) Russian Academy of Sciences, Moscow, Russia
 }
\date{Received April 21, 2005; accepted sometime 2005}

\abstract{We present a new model of the heliospheric interface -
the region of the solar wind interaction with the local
interstellar medium. This new model performs a multi-component
treatment of charged particles in the heliosphere. All charged
particles are divided into several co-moving types. The coldest
type, with parameters typical of original solar wind protons, is
considered in the framework of fluid approximation. The hot pickup
proton components created from interstellar H atoms and
heliospheric ENAs by charge exchange, electron impact ionization
and photoionization are treated kinetically. The charged
components are considered self-consistently with interstellar H
atoms, which are described kinetically as well. To solve the
kinetic equation for H atoms we use the Monte Carlo method with
splitting of trajectories, which allows us 1) to reduce
statistical uncertainties allowing correct interpretation of
observational data, 2) to separate all H atoms in the heliosphere
into several populations depending on the place of their birth and
on the type of parent protons.
}
 \maketitle

\section{Introduction}
The heliospheric interface is formed by interaction of the solar
wind with the local interstellar medium (LISM). This is a complex
region where the magnetized solar wind and interstellar plasmas,
interstellar atoms, galactic and anomalous cosmic rays and pickup
ions all play prominent roles. Our Moscow group has developed
models of the heliospheric interface, which follow the Baranov \&
Malama (1993) kinetic-continuum approach and take into account
effects of the solar cycle (Izmodenov et al., 2005a), galactic
cosmic rays (Myasnikov et al., 2000), interstellar helium ions and
solar wind alpha particles (Izmodenov et al., 2003), the
interstellar magnetic field (Alexashov et al., 2000; Izmodenov et
al, 2005b), and anomalous cosmic rays (Alexashov et al, 2004; for
review, see Izmodenov, 2004). However, all of these models assume
immediate assimilation of pickup protons into the solar wind
plasma and consider the mixture of solar wind and pickup protons
as a single component.

Most other models also make this assumption (see for review, Zank
1999). However, it is clear from observations (e.g. Gloeckler and
Geiss, 2004) that the pickups are thermally decoupled from the
solar wind protons and should be considered as a separate
population. Moreover, measured spectra of pickup ions show that
their velocity distributions are not Maxwellian. Therefore, a
kinetic approach should be used for this component. Theoretical
kinetic models of pickup ion transport, stochastic acceleration
and evolution of their velocity distribution function are  now
developed (Fisk, 1976; Isenberg, 1987; Bogdan et al., 1991;
Fichtner et al., 1996; Chalov et al., 1997; le Roux \& Ptuskin,
1998). However, mostly these models are 1) restricted by the
supersonic solar wind region; 2) do not consider the back reaction
of pickup protons on the solar wind flow pattern, i.e. pickup
protons are considered as test particles. Chalov et al. (2003,
2004a) have studied properties of pickup proton spectra in the
inner heliosheath, but in its upwind part only. Several
self-consistent multi-component models (Isenberg, 1986; Fahr et
al. 2000; Wang \& Richardson, 2001) were considered, however
pickup ions in these models were treated in the fluid
approximation that does not allowed to study kinetic effects.

 In this paper we present our new kinetic-continuum model of the
heliospheric interface.The new model retains the main advantage of
our previous models, that is a rigorous kinetic description of the
interstellar H atom component. In addition, it considers pickup
protons as a separate kinetic component.

\section{Model}

Since the mean free path of H atoms, which is mainly determined by
the charge exchange reaction with protons, is comparable with the
characteristic size of the heliosphere, their dynamics is governed
by the kinetic equation for the velocity distribution function
$f_{\rm H}(\vec{r}, \vec{v}_{\rm H}, t)$:
\[
\frac{\partial f_{\rm H}}{\partial t}+ \vec{v}_{\rm H} \cdot
\frac{\partial f_{\rm H}}{\partial \vec{r} } +
\frac{\vec{F}}{m_{\rm H}} \cdot \frac{\partial f_{\rm H}}{\partial
\vec{v}_{\rm H} } = - ( \nu_{ph} + \nu_{\rm impact} ) f_{\rm H} (
\vec{r}, \vec{v}_{\rm H} )
\]
\begin{equation} \label{eqBoltz}
\noindent
 \nonumber - f_{\rm H} \cdot \sum_{i=p,pui} \int |
\vec{v}_{\rm H} - \vec{v}_i | \sigma^{\rm HP}_{ex} f_i (\vec{r},
\vec{v}_i) d \vec{v}_i
\end{equation}
\[
+ \sum_{i=p,pui} f_i(\vec{r}, \vec{v}_{\rm H}) \int | \vec{v}_{\rm
H}^* - \vec{v}_{\rm H} | \sigma^{\rm HP}_{ex} f_{\rm H} (\vec{r},
\vec{v}_{\rm H}^* ) d \vec{v}_{\rm H}^* .
\]
Here $ f_p( \vec{r}, \vec{v}_p ) $ and $ f_{pui}( \vec{r},
\vec{v}_{pui} ) $ are the local distribution functions of protons
and pickup protons; $ \vec{v}_p $, $ \vec{v}_{pui}$  and $
\vec{v}_{\rm H}$ are the individual proton, pickup proton, and
H-atom velocities in the heliocentric rest frame, respectively; $
\sigma^{\rm HP}_{ex} $ is the charge exchange cross section of an
H atom with a proton; $ \nu_{ph} $ is the photoionization rate; $
m_{\rm H}$ is the atomic mass; $ \nu_{\rm impact} $ is the
electron impact ionization rate; and $ \vec{F} $ is the sum of the
solar gravitational force and the solar radiation pressure force.

We consider all plasma components (electrons, protons, pickup
protons, interstellar helium ions and solar wind alpha particles)
as media co-moving with bulk velocity $\vec{u}$. The plasma is
quasi-neutral, i.e. $n_e = n_p+ n_{He^+}$ for the interstellar
plasma and $n_e = n_p+n_{pui}+2 n_{He^{++}}$ for the solar wind.
For simplicity we ignore the magnetic field. While the interaction
of interstellar H atoms with protons by charge exchange is
important, this process is negligible for helium due to small
cross section. The system of governing equations for the sum of
all ionized components is:
\begin{eqnarray}
\frac{\partial \rho }{\partial t}+ div (\rho \vec{u})=q_1,
\nonumber \\
\frac{\partial \rho \vec{u}}{\partial t}+ div (\rho
\vec{u}\vec{u}+ p \hat{I}) = \vec{q}_2 \\
 \frac{\partial
E}{\partial t} + div ([E+p]\vec{u}) = q_3 +q_{3,e} \nonumber
\end{eqnarray}
Here $\rho = \rho_p+\rho_e+ \rho_{He}+ \rho_{pui} $ is the total
density of the ionized component, $p = p_p+p_e+p_{pui}+p_{He}$ is
the total pressure of the ionized component, $E=\rho (\varepsilon+
\vec{u}^2/2)$ is the total energy per unit volume, $\varepsilon =
p/(\gamma-1)\rho$ is the specific internal energy.

The expressions for the sources are following:
\[
q_1= m_{\rm p} n_{\rm H} \cdot (\nu_{\rm ph} + \nu_{\rm impact}),
\qquad n_{\rm H}= \int f_{\rm H}({\vec{v}}_{\rm H}) d
{\vec{v}}_{\rm H},
\]
\[
\vec{q}_2 =  \int m_{\rm p} (\nu_{\rm ph}+\nu_{\rm impact})
\vec{v}_{\rm H} f_{\rm H}(\vec{v}_{\rm H}) d {\vec{v}_{\rm H}} +
\]
\[
\int \int m_{\rm p} v_{\rm {rel}} \sigma_{ex}^{HP}(v_{\rm {rel}})
(\vec{v}_{\rm H} - \vec{v}) f_{\rm H}(\vec{v}_{\rm H})
\sum_{i=p,pui}f_i({\vec{v}}) d \vec{v}_{\rm H} d \vec{v},
\nonumber
\]
\[ q_3 =  \int m_{\rm p}
(\nu_{\rm ph}+\nu_{\rm impact})\frac{\vec{v}_{\rm H}^2}{2} f_{\rm
H}(\vec{v}_{\rm H}) d \vec{v}_{\rm H} + \frac{1}{2} \int \int
m_{\rm p} v_{\rm {rel}} \cdot
\]
\[
\cdot  \sigma^{\rm HP}_{ex}(v_{\rm {rel}}) (\vec{v}_{\rm H}^2 -
\vec{v}^2) f_{\rm H}(\vec{v}_{\rm H}) \sum_{i=p,pui}f_i(\vec{v}) d
\vec{v}_{\rm H} d \vec{v},  \nonumber
\]
\[
q_{3,e} = n_{\rm H} (\nu_{\rm ph}E_{\rm ph} - \nu_{\rm
impact}E_{\rm ion}) , \nonumber
\]
$v_{\rm {rel}}=|\vec{v}_{\rm H} - \vec{v}|$ is the relative
velocity of an atom and a proton, $E_{\rm ph}$ is the mean
photoionization energy (4.8 eV), and $E_{\rm ion}$ is the
ionization potential of H atoms (13.6 eV).
\begin{figure}
\includegraphics[width=8cm]{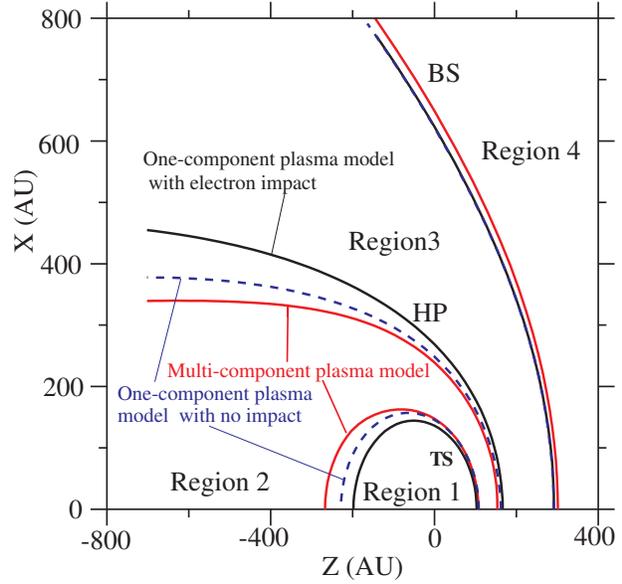}
\caption{The termination shock, heliopause and bow shock shown for
three models of the heliospheric interface: 1) new multi-component
model, 2) Baranov \& Malama model, 3) Baranov \& Malama model with
no electron impact. \label{fig1}}
\end{figure}

 The system of the equations for the velocity distribution
function of H-atoms (Eq.~(1)) and for mass, momentum and energy
conservation for the total ionized component (Eqs.~(2)) is not
self-consistent, since it includes the velocity distribution
function of pickup protons. At the present time there are many
observational evidences (e.g. Gloeckler et al. 1993, Gloeckler
1996, Gloeckler \& Geiss 1998) and theoretical estimates (e.g.
Isenberg 1986), which clearly show that pickup ions constitute a
separate and very hot population in the solar wind. Even though
some energy transfer from the pickup ions to solar wind protons is
now theoretically admitted in order to explain the observed
heating of the outer solar wind (Smith et al. 2001; Isenberg et
al. 2003, Richardson \& Smith 2003, Chashei et al. 2003, Chalov et
al. 2004b), it constitutes not more than 5\% of the pickup ion
energy. The observations show also that the velocity distribution
function can be considered as isotropic (fast pitch-angle
scattering) except some short periods in the inner heliosphere
when the interplanetary magnetic field is almost radial.

So we assume here that the velocity distribution of pickup protons
in the solar wind rest frame is isotropic, and it is determined
through the velocity distribution function in the heliocentric
coordinate system by the expression:
\begin{equation}
f_{pui}^*(\vec{r}, w) = \frac{1}{4 \pi}\int\int f_{pui}(\vec{r},
\vec{v}) sin \theta d \theta d \phi .
\end{equation}
Here $\vec{v} = \vec{u} + \vec{w}$, $\vec{v}$ and $\vec{u}$ are
the velocity of a pickup proton and bulk velocity of the plasma
component in the heliocentric coordinate system, $\vec{w}$ is the
velocity of the pickup proton in the solar wind rest frame, and
($w$, $\theta$, $\phi$) are coordinates of $\vec{w}$ in the
spherical coordinate system. The equation for $f_{pui}^*(\vec{r},
w)$ can be written in the following general form taking into
account velocity diffusion but ignoring spatial diffusion, which
is unimportant at the energies under consideration:
\begin{equation}
\frac{\partial f_{pui}^*}{\partial t} + \vec{u} \cdot
\frac{\partial f_{pui}^*}{\partial \vec{r}} =
\frac{1}{w^2}\frac{\partial}{\partial w}\left( w^2 D
\frac{\partial f_{pui}^*}{\partial w} \right)+
\end{equation}
\[
 +
\frac{w}{3} \frac{\partial f_{pui}^*}{\partial w}
div(\vec{u})+S(\vec{r},w) ,
\]
where $D(\vec{r},w) $ is the velocity diffusion coefficient. The
source term $S(\vec{r}; w)$ can be written as
\begin{equation}
S(\vec{r}; w) = \frac{1}{4 \pi} \int \int
\nu_{ion}(\vec{w})f_H(\vec{r}, \vec{w}+\vec{u}) sin \theta d
\theta d \phi
\end{equation}
\[
- \frac{1}{4 \pi} \int \int f_{pui}^*(\vec{r}, w) \nu_H(\vec{w})
sin \theta d \theta d \phi .
\]
In equation (5) $\nu_{ion}$ and $\nu_H$ are ionization rates:
\[
\nu_{ion} = \sum_{i=pui,p} \int\int\int f_i^*(\vec{r}, w_i) |
\vec{w}_i - \vec{w}| \times
\]
\[
\sigma^{\rm HP}_{ex}(| \vec{w}_i - \vec{w}|)w_i^2 dw_i sin \theta
_{i} d \theta _{i} d \phi _{i}+ \nu_{\rm ph}+ \nu_{\rm impact} ,
\]
\[
\nu_{H} =  \int f_H(\vec{r}, \vec{v}_H) | \vec{w} + \vec{u} -
\vec{v}_H| \sigma^{\rm HP}_{ex}(| \vec{w} + \vec{u} - \vec{v}_H|)
\, d\vec{v}_H.
\]
Although only stationary solutions of Eq.~(4) will be sought here,
we prefer to keep the first term in the equation to show its
general mathematical structure. The effective thermal pressure of
the pickup ion component is determined by
\begin{equation}
p_{pui}=\frac{4 \pi}{3}\int m_p w^2 f_{pui}^*(\vec{r}, w) w^2 dw.
\end{equation}

We solve the continuity equations for He$^+$ in the interstellar
medium and for alpha particles in the solar wind. Then proton
number density can be calculated as $n_p = (\rho - m_{He}
n_{He})/m_p -n_{pui}$.  Here $n_{He}$ denotes the He$^+$ number
density in the interstellar medium, and He$^{++}$ the number
density in the solar wind.

In addition to the system of equations (1), (2), (4) we solve the
heat transfer equation for the electron component:
\begin{eqnarray}
\frac{\partial \rho_e \varepsilon_e}{\partial t} + div ([\rho_e
\varepsilon_e]\vec{u}) =
\end{eqnarray}
\[
-p_e div \vec{u}+ q_{3,e}+ Q_{e,p}+Q_{e,pui}.
\]
Here $\varepsilon_e = p_e/(\gamma-1)\rho_e $ is the specific
internal energy of the electron component, $Q_{e,p}$ and
$Q_{e,pui}$ are the energy exchange terms of electrons with
protons and pickup ions, respectively.

To complete the formulation of the problem we need to specify: a)
the diffusion coefficient $D(\vec{r},w)$, b) exchange terms
$Q_{e,p}$, $Q_{e,pui}$, c) the behavior of pickup protons and
electrons at the termination shock. In principle, our model allows
us to make any assumptions and verify any hypothesis regarding
these parameters. Moreover, the diffusion coefficient
$D(\vec{r},w)$ depends on the level of solar wind turbulence, and
equations describing the production (say, by pickups) and
evolution of the turbulence need to be added to equations (1)-(7).
This work is still in progress and will be described elsewhere.

In this paper we consider as simple a model as possible. We adopt
$D=0$. While velocity diffusion is not taken into account in the
paper, suprathermal tails in the velocity distributions of pickup
protons are formed as will be shown below.

It is believed that the thickness of the termination shock ramp
lies in the range from the electron inertial length up to the ion
inertial length: $c/\omega_{\mathrm e} \leq L_{\mathrm ramp} \leq
c/\omega_{\mathrm p}$, where $\omega_{\mathrm e,p} = \left( 4\pi
n_{\mathrm e,p} e^2/m_{\mathrm e,p}\right) ^{1/2}$ are the
electron and proton plasma frequencies (see discussion in Chalov
2005). Since this thickness is less than the gyro-radius of a
typical pickup proton in front of the termination shock at least
by a factor of 10, the shock is considered here as a
discontinuity. In this case the magnetic moment of a pickup ion
after interaction with a perpendicular or quasi-perpendicular
shock is the same as it was before the interaction (Toptygin 1980,
Terasawa 1979). The only requirement is that the mean free path of
the ions is larger than their gyro-radius (weak scattering). For
perpendicular or near perpendicular parts of the termination shock
conservation of the magnetic moment leads to the following jump
condition at the shock (Fahr \& Lay 2000):
\begin{equation}
f_{2,pui}^*(\vec{r},w)= C^{-1/2} f_{1,pui}^*(\vec{r},w/\sqrt{C})
\, \label{p1}
\end{equation}
 where $C= \rho_2/\rho_1$ is the shock compression. Although
the termination shock can be considered as perpendicular at their
nose and tail parts only (Chalov \& Fahr 1996, 2000; Chalov 2005),
Nevertheless, we assume here for the sake of simplicity that
Eq.~(\ref{p1}) is valid everywhere at the shock. It should be
emphasized that we are forced to adopt this condition in our
axisymmetric model because taking into account real geometry of
the large-scale magnetic field near the shock requires one to take
into account 1) the three-dimensional structure of the interface,
2) reflection of pickup ions at the termination shock due to
abrupt change in the magnetic field strength and direction. While
the reflection is a very important process to inject ions into
anomalous cosmic rays, we will consider this complication of our
present model in future work. Note that the concept of magnetic
moment conservation is only one of the possible scenarios of the
behavior of pickup ions at the termination shock. Other
possibilities will also be considered in the future.

We assume also that $Q_{e,pui}=0$ and $Q_{e,p}$ is such that
\[
T_e=T_p = T_{He^{++}}
\]
everywhere in the solar wind. Because of this assumption, there is
no need to solve Es.~(7). Nevertheless, we solve this equation in
order to check numerical accuracy of our solution. Later, we plan
to explore models with more realistic $D$ ,$Q_{e,pui}$ and
$Q_{e,p}$ and different microscopic theories can be tested.

The boundary conditions for the charged component are determined
by the solar wind parameters at the Earth's orbit and by
parameters in the undisturbed LIC. At the Earth's orbit it is
assumed that proton number density is n$_{p,E}$=n$_{e,E}$=7.72
cm$^{-3}$, bulk velocity is $u_E$=447.5 km/s and ratio of alpha
particle to proton is 4.6 \%. These values  were obtained by
averaging the OMNI 2 solar wind data over two last solar cycles.
 The velocity and temperature of the pristine
interstellar medium were recently determined from the
consolidation of all available experimental data
(M$\mathrm{\ddot{o}}$bius et al. 2004, Witte 2004, Gloeckler 2004,
Lallement 2004a,b). We adopt in this paper V$_{LIC}$= 26.4 km/s
and T$_{LIC}$=6527 K.

 For the local interstellar H atom,
proton and helium ion number densities we assume n$_{H,LIC}$ =
0.18 cm$^{-3}$, n$_{p,LIC}$=0.06 cm$^{-3}$ and
n$_{He^{+},LIC}$=0.009 cm$^{-3}$ , respectively (for
argumentation, see, e.g., Izmodenov et al. 2003, 2004). The
velocity distribution of interstellar atoms is assumed to be
Maxwellian in the unperturbed LIC. For the plasma component at the
outer boundary in the tail we used soft outflow boundary
conditions. For the details of the computations in the tail
direction see Izmodenov \& Alexashov (2003), Alexashov et al.
(2004b).

To solve the system of governing Euler equations for the plasma
component, the second order finite volume Godunov type numerical
method was used (Godunov et al. 1979; Hirsch 1988). To increase
the resolution properties of the Godunov scheme, a piecewise
linear distribution of the parameters inside each cell of the grid
is introduced. To achieve the TVD property of the scheme the
minmod slope limiter function is employed (Hirsch 1988). We used
an adaptive grid as in Malama (1991) that fits the termination
shock, the heliopause and the bow shock. The kinetic equation (1)
was solved by the Monte-Carlo method with splitting of
trajectories following Malama (1991). The Fokker-Planck type
equation (4) for the pickup proton velocity distribution function
is solved
 by calculating statistically relevant numbers of stochastic particle trajectories
 (Chalov et al., 1995). To get a self-consistent solution of the plasma Euler equations
(2), kinetic equation (1) and Fokker-Planck type eq. (4) we used
the method of global iterations suggested by Baranov et al.
(1991).

\begin{table}
  \centering
  \caption{Description of introduced types of protons and populations of H atoms.}\label{tab1}
\begin{tabular}{ll}
  \hline
Type number & Proton type description\\
  \hline
  0 &  'cold proton type' consisting of:  \\
    & original solar wind protons +  protons  \\
    &  created in region 1 from atoms of  \\
    &  population 1.0 + protons created \\
    &  in region 2 from atoms of populations  \\
    &  2.0, 3, 4 \\
  1 & pickup protons created in region 1 from \\
    & atoms of populations 1.1, 2.0, 3, 4 \\
  2 & pickup protons created in region 1 from \\
    & atoms of populations 1.2, 2.1-2.4 \\
  3 & pickup protons created in region 2 from \\
    & atoms of populations 1.0, 1.1 \\
  4 & pickup protons created in region 2 from \\
    & atoms of populations 1.2, 2.1-2.4 \\
      \hline
Population number &  H atoms created in:\\
  \hline
  1.0 &  region 1 from protons of type 0 \\
  1.1 &  region 1 from pickup protons of type 1 \\
  1.2 &  region 1 from pickup protons of type 2 \\
  2.0 & region 2 from  protons of type 0 \\
  2.1 & region 2 from pickup protons of type 1 \\
  2.2 & region 2 from pickup protons of type 2 \\
  2.3 & region 2 from pickup protons of type 3 \\
  2.4 & region 2 from pickup protons of type 4 \\
  3 & secondary interstellar atoms (as previously) \\
  4 & primary interstellar atoms (as previously) \\
    \hline
\end{tabular}
\end{table}

\begin{figure}
\includegraphics[width=\linewidth]{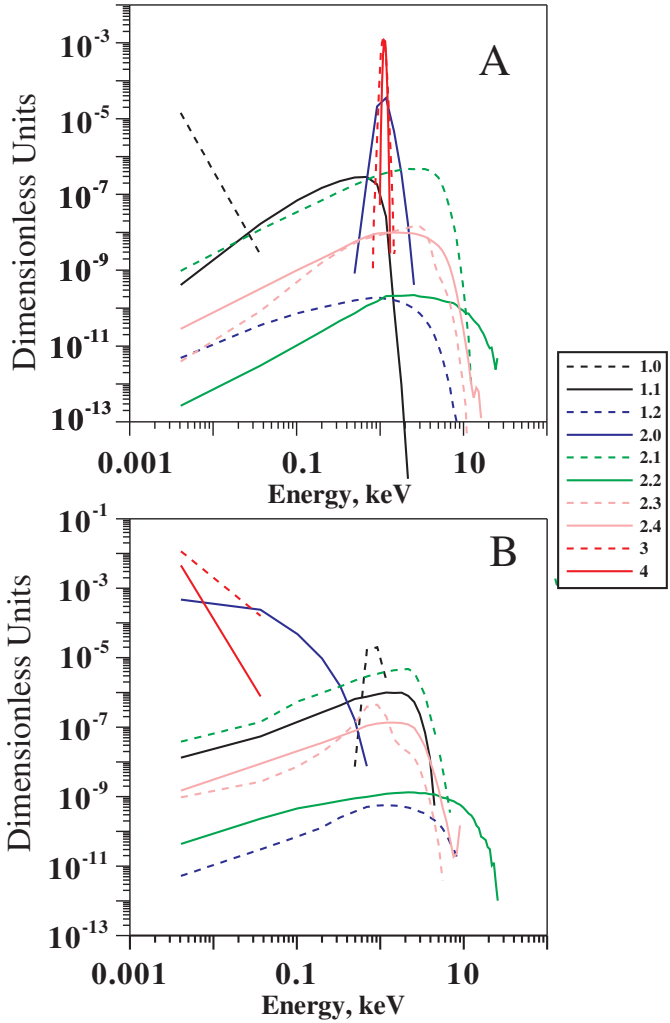}
\caption{The source term S (Eq.~(4)) from different populations of
H atoms as a function of energy shown in the supersonic solar wind
at 5 AU (A) and in the inner heliosheath (B). \label{fig2}}
\end{figure}

\section{Results of numerical modeling}

Figures 1-6 present the main results obtained in the frame of our
new multi-component model described in the previous section. The
shapes and locations of the termination shock (TS), heliopause
(HP) and bow shock (BS) are shown in Fig. 1. For the purposes of
comparison the positions of the TS, HP, and BS are also shown in
the case when pickup and solar wind protons are treated as a
single fluid. Later we refer to this model as the Baranov-Malama
(B\&M) model. Two different cases obtained with the B\&M model are
shown. In the first case ionization by electron impact is taken
into account, while this effect is omitted in the second case. The
only (but essential) difference between the B\&M model and our new
model, considered in this paper, is that the latter model treats
pickup protons as a separate kinetic component. As seen from
Figure 1 the differences in the locations of the TS, HP, and BS
predicted by the new and B\&M models are not very large in the
upwind direction. The TS is 5 AU further away from the Sun in the
new model  compared to B\&M models. The HP is 12 AU closer. The
effect is much more pronounced in the downwind direction where the
TS shifts outward from the Sun by $\sim$70 AU in the new model.
Therefore, the inner heliosheath region is thinner in the new
model compared to the B\&M model. This effect is partially
connected with lower temperature of electrons and, therefore, with
a smaller electron impact ionization rate in this region. Indeed,
new pickup protons created by electron impact deposit additional
energy and, therefore, pressure in the region of their origin,
i.e. in the inner heliosheath. The additional pressure pushes the
heliopause outward and the TS toward the Sun. Even though our
multi-component model takes into account ionization by electron
impact, this is not as efficient as in one-fluid models (like
B\&M) due to the lower electron temperature in the heliosheath.
Excessively high electron temperatures which are predicted by the
one-fluid models in the outer heliosphere are connected with the
physically unjustified assumption of the immediate assimilation of
pickup protons into the solar wind plasma.

However, the HP is closer to the Sun and the TS is further from
the Sun in the new multi-component model  even in the case when
electron impact ionization is not taken into account. This is
because the solar wind protons and pickup protons are treated in
the new multi-component model as two separate components. Indeed,
hot energetic atoms (ENAs), which are produced in the heliosheath
by charge exchange of interstellar H atoms with both the solar
wind protons and pickup protons heated by the TS, escape from the
inner heliosheath easily due to their large mean free paths. These
ENAs remove (thermal) energy from the plasma of the inner
heliosheath and transfer the energy to other regions of the
interface (e.g., into the outer heliosheath). In the case of the
new model there are  two parenting proton components for the ENAs
- the original solar protons and pickup protons. In the B\&M model
these two components are mixed to one. As a result, the ENAs
remove energy from the inner heliosheath more efficiently in the
case of the multi-component model than the B\&M model not. Similar
effect was observed for multi-fluid models of H atoms in the
heliospheric interface described in detail by Alexashov \&
Izmodenov (2005).

 To gain a better insight into the results of the new model
and its potential possibilities to predict and interpret
observational data, we divide heliospheric protons (original solar
and pickup protons) into five types, and H atoms into ten
populations described in Table 1. The first index in the notation
of an H-atom population is the number of the region, where the
population was created, i.e. populations 1.0-1.2 are the H atoms
created in the supersonic solar wind (region 1, see Fig.1),
populations 2.0-2.4 are the H atoms created in the inner
heliosheath (region 2), populations 3 and 4 are secondary and
primary interstellar atoms. Definitions of the two last
populations are the same as in the B\&M model. The second index
denotes the parent charged particles (protons), i.e. from 0 to 4.

 Original solar wind protons are denoted
as type 0. Protons of this type are cold compared to the normal
pickup protons in the solar wind. The pickup protons, which have
characteristics close to the original solar wind protons, are also
added to type 0. These are pickup protons created in the
supersonic solar wind (region 1) from H atoms of population 1.0
(this population forms a so-called neutral solar wind, e.g.
Bleszynski et al. 1992) and pickup protons created in the inner
heliosheath (region 2) from H atoms of populations 2.0, 3, and 4.
The type 0 is formed in such a way that 1) its thermal pressure is
much less than the dynamic pressure everywhere in the heliosphere
and, therefore, unimportant; 2) we are not interested in details
of the velocity distribution of this type of protons and assume
that it is Maxwellian. The rest of the pickup protons is divided
into four types: two types are those pickup protons that are
created in region 1 (supersonic solar wind), and the others are
pickup protons created in region 2 (inner heliosheath). In each
region of birth we separate pickup protons into two additional
types depending on their energy  (more precisely,  parent atoms).
For instance, type 1 is the ordinary pickup proton population
which is created in the supersonic solar wind from primary and
secondary interstellar atoms  and then convected in the inner
heliosheath. Type 2 is also created in the supersonic solar wind
but, in distinction to type 1, from energetic atoms. Among pickup
protons created in the inner heliosheath type 4 is more energetic
than type 3. Thus, two types of pickup protons (1 and 2) exist in
the supersonic solar wind and four (1-4) in the inner heliosheath,
from which types 2 and 4 are more energetic than 1 and 3. A more
detailed description of the properties of pickup protons of
different types will be given below (Figs. 3-4).

\begin{figure}
\includegraphics[width=8cm]{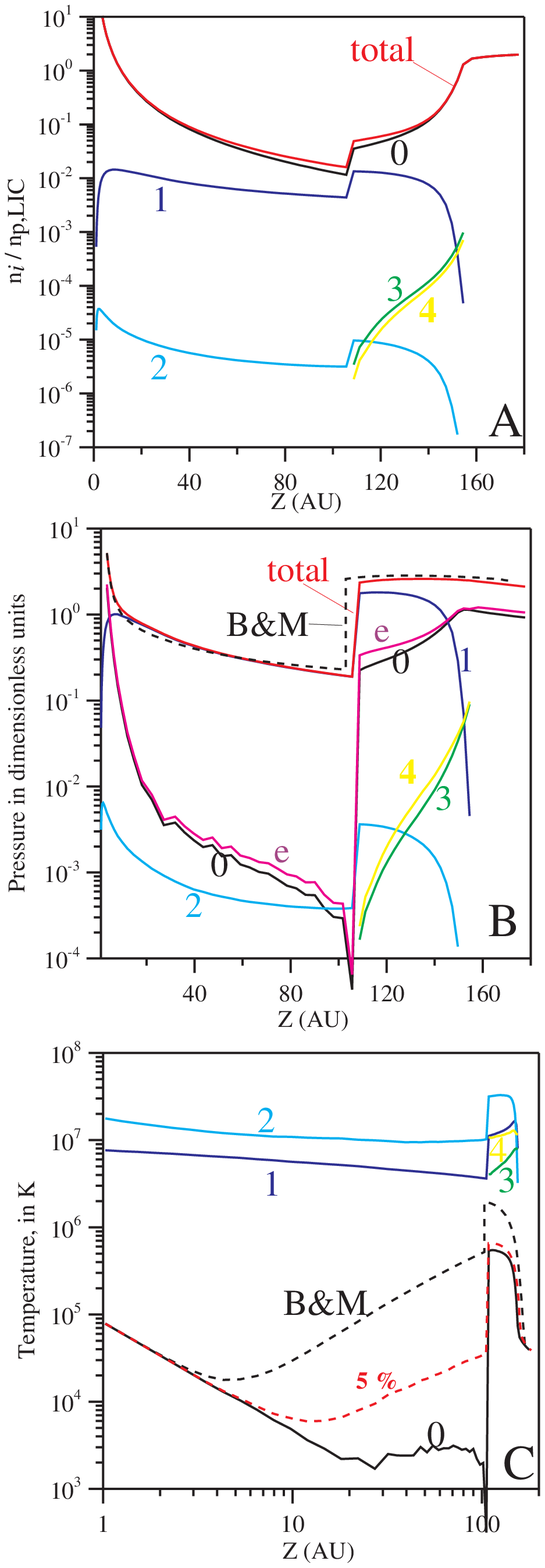}
\caption{Number densities (A), thermal pressure (B) and
temperature (C) of different types of protons. Curves are labelled
with proton types. \label{fig3}}
\end{figure}

One should keep in mind that at any place in the heliosphere real
pickup protons constitute a full distribution and they are not
divided into different types as we introduce here. However, and it
is very important, the pickup protons have a rather broad energy
spread, and, as we show below, particles with different places of
birth and parent atoms prevail in a particular energy range
according to our separation into the different types. The same is
valid for H atoms. Thus one of the main advantages of our new
model is the theoretical possibility to predict solar wind plasma
properties in the outer parts of the heliosphere (including the
inner heliosheath) through observations of pickup protons and
hydrogen atoms in different energy ranges. Note that even without
this rather complicated separation of particles into different
types and populations but in the case when all pickup protons are
treated as a distinct kinetic component (the simplest version of
our model), the plasma flow pattern, positions and shapes of the
TS, HP and BS are essentially the same.

As an useful illustration of the aforesaid, we present the
calculated source term $S$ (see Eq. (4)) of pickup protons in the
supersonic solar wind and in the heliosheath (Fig. 2). It is
apparent that relative contributions of different H-atom
populations into pickup protons are essentially different. In the
supersonic solar wind (Fig. 2A) narrow peaks  near 1 keV are
created by populations 2.0, 3 and 4. These populations are seeds
for the major type of pickup protons, which we denote as type 1.
Pickup protons created from population 1.1 are also added to type
1 due to the lack of high energy tails in their distributions.
Pickup protons  created from H atoms of populations 1.2, 2.1-2.4
form type 2, which is more energetic compared to type 1 (see also
argumentation above). A similar discussion can be applied to
pickup protons created in the inner heliosheath.
\begin{figure*}
\includegraphics[width=\linewidth]{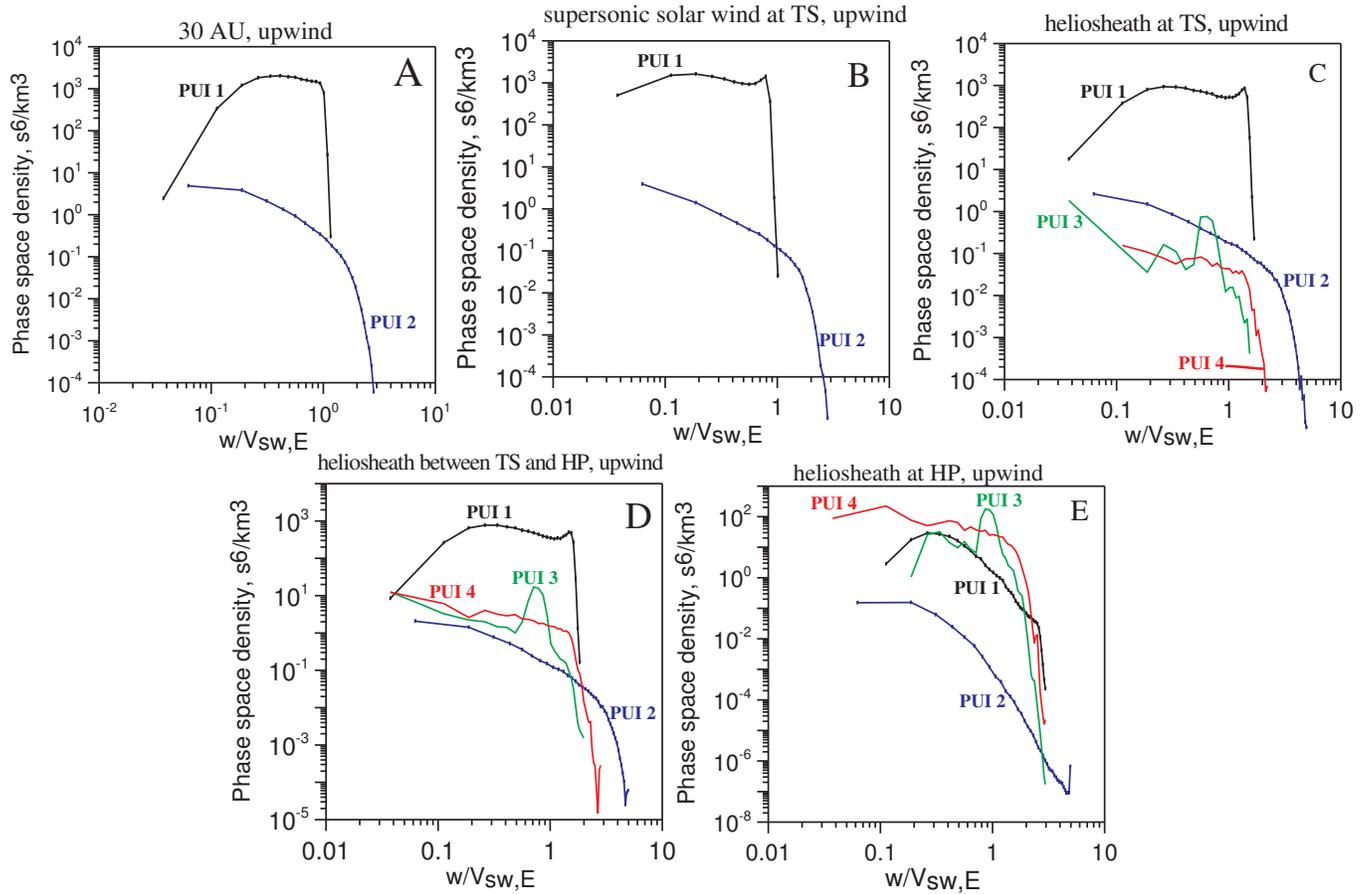}
\caption{ Phase space densities of different types of protons in
the supersonic solar wind at 30 AU (A), upstream (B) and
downstream (C) of the termination shock, inside the inner
heliosheath (D) and at the heliopause (E). All curves are shown
for the upwind direction.\label{fig4}}
\end{figure*}
It is important to underline here  again that the main results of
our model do not depend on the way we divide H atoms and pickup
ions into populations and types. Such a division has two principal
goals: 1) to have a clearer insight into the origin and nature of
the pickup ions measured by SWICS/Ulysses and ACE (Gloeckler \&
Geiss 2004) and ENAs that will be measured in the near future
(McComas et al. 2004), and 2) to obtain better statistics in our
Monte Carlo method with splitting of trajectories (Malama 1991)
when we calculate high energy tails in the distributions of pickup
protons and H atoms that are several orders of magnitude lower
than the bulk of particles.

\begin{figure*}
\includegraphics[width=\linewidth]{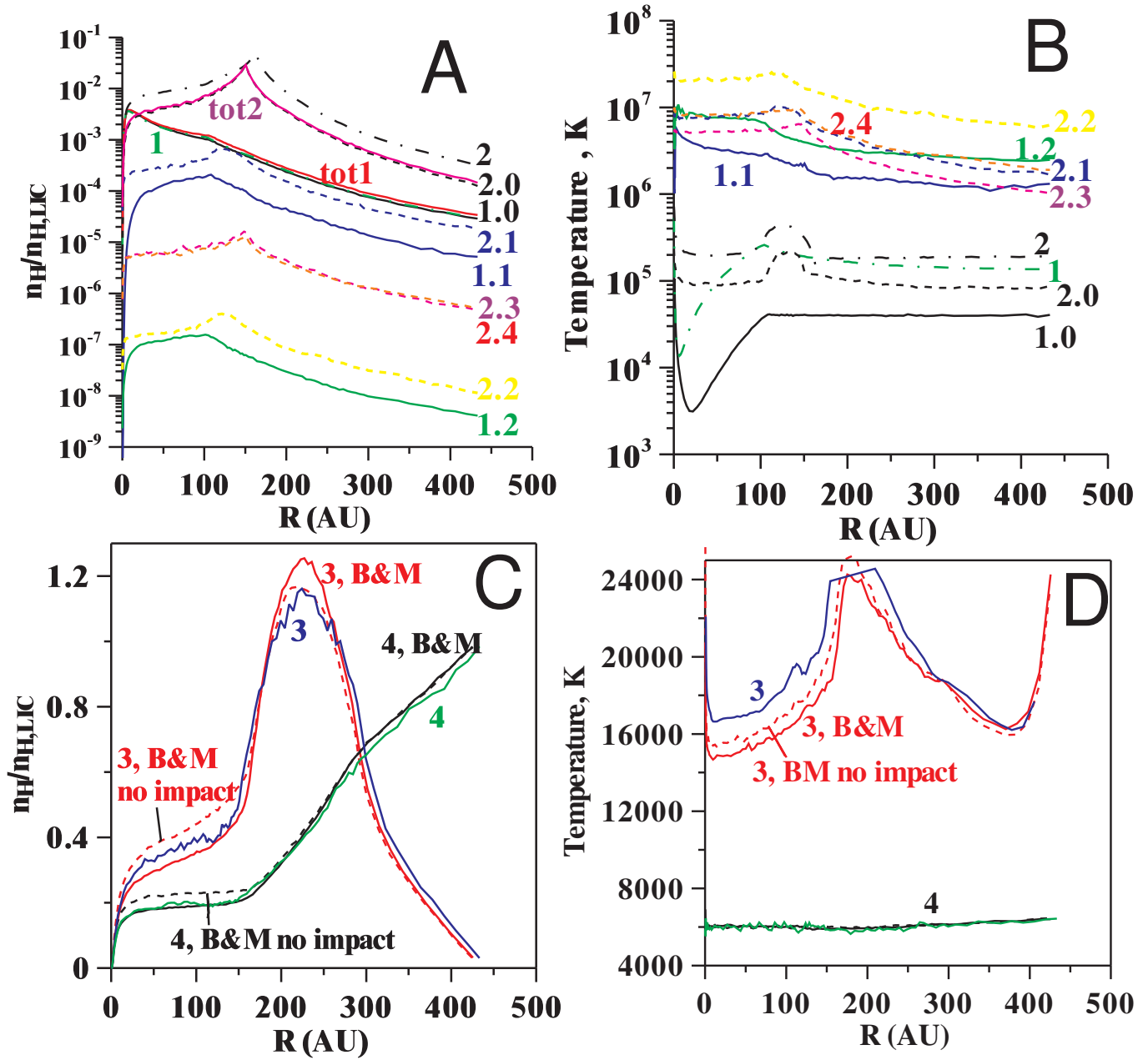}
\caption{Number densities of H-atom populations 1.0-1.2, 2.0-2.4
(A) and populations 3,4 (C), and temperatures of these populations
(B and D, respectively) as functions of heliocentric distance.
Numeration of curves corresponds to numeration of the populations.
Number densities and temperatures obtained with the Baranov and
Malama (1993) model denoted as B\&M and "B\&M no impact" are shown
for comparison. The sum of the number densities of populations
1.0,1.1 and 1.2 are shown as curve 'tot1'. The sum of the number
densities of populations 2.0-2.4 are shown as curve 'tot2'.
\label{fig5}}
\end{figure*}

Number densities, pressures and temperatures for the introduced
types of charged particles are shown in Fig. 3. It is seen (Fig.
3A) that the protons of type 0 (recall that these protons are
mainly of solar origin) dominate by number density everywhere in
the heliosphere, while type 1 of pickup protons makes up to 20 \%
of the total number density in the vicinity of the TS. Approaching
the HP the number densities of types 1 and 2 decrease, since in
accordance with our notation these types are created in the
supersonic solar wind only. When the pickup protons are convected
in the regions behind the TS, types 1 and 2 experience losses due
to charge exchange with H atoms. New pickup protons created in the
inner heliosheath as a result of this reaction have different
properties than the original pickup protons (see below), and we
assign them to types 3 and 4. The number densities of these types
increase towards the HP.

Although the solar wind protons (type 0) are in excess in the
number density, nevertheless pickup protons are generally much
hotter (see Fig. 3C) and as a consequence the thermal pressure of
type 1 dominates in the outer parts of the supersonic solar wind
and in the inner heliosheath. Downstream of the TS this pressure
is almost an order of magnitude larger than the pressure of type
0. As would be expected, the temperatures of types 2 and 4 are
much larger than those of types 1 and 3, respectively. The
temperature of the solar wind protons of type 0 decreases
adiabatically up to 20 AU. Then the temperature becomes so low
that the energy transferred to the solar wind by photoelectrons
becomes non-negligible.  This effect results in the formation of a
plateau in the spatial distribution of the proton temperature in
the region from 20 AU up to the TS.  Figure 3C presents  the
proton temperature calculated with the B\&M model for comparison.
A highly unrealistic increase in the temperature is connected to
the unrealistic assumption of immediate assimilation of pickup
protons into the solar wind.  Our new model is free of this
assumption but it allows us to take into account some energy
transfer between pickup and solar wind protons. The curve denoted
as "5\%" in Fig. 3C shows the results of calculations, where we
simply assume that 5 \% of the thermal energy of pickup protons is
transferred to the solar wind protons (to type 0).
 For pickup
protons it is assumed that initial  velocities (in the solar wind
rest frame) of newly injected pickup protons are $\sim$2.5\%
smaller than the solar wind bulk velocity.
 The
curve denoted as "5\%" is qualitatively very similar to the
Voyager 2 observations
 which clearly show some increase in the proton temperature in
the outer heliosphere. More prominent mechanisms of energy
transfer between pickup and solar wind protons  based on real
microphysical background will be included in the model in the
future.

Velocity distribution functions (in the solar wind rest frame) of
the four types of pickup protons are shown in Fig. 4 for different
heliocentric distances in the upwind direction. All distributions
are presented as functions of the dimensionless speed $w/V_{\rm
SW,E}$, where $V_{\rm SW,E}$ is the solar wind speed at the
Earth's orbit. In the supersonic solar wind   (Fig. 4A,B)  type 1
is dominant at energies below   about  1 keV   ($w < V_{\rm
SW,E}$), while the more energetic type 2 is dominant for energies
above 1 keV ($w > V_{\rm SW,E}$). As was shown for the first time
by Chalov \& Fahr (2003), this energetic type of pickup protons
(secondary pickup protons) created in the supersonic solar wind
from energetic hydrogen atoms can form the quite-time suprathermal
tails  observed by SWICS/Ulysses and ACE instruments (Gloeckler
1996, Gloeckler \& Geiss 1998). Downstream of the TS up to the
heliopause (Fig. 4C-E) the high-energy tails are more pronounced.
The high-energy pickup ions form an energetic population of H
atoms  known as ENAs (e.g. Gruntman et al. 2001).

 As we have discussed above and as is seen from Fig. 4, the
pickup protons of type 1 prevail throughout the heliosphere except
a region near the HP. In the supersonic solar wind the velocity
distributions of these pickup protons are close to the
distributions obtained by Vasyliunas \& Siscoe (1976) who also
ignored velocity diffusion. However,  our distributions are
different, since in our model 1) the solar wind speed varies with
the distance from the Sun, 2) the spatial behavior of the
ionization rate is more complicated, 3) thermal velocities of H
atoms are taken into account, 4) the set of parent atoms is more
varied (see Table 1). Note that we are forced to calculate the
velocity distribution functions in the supersonic solar wind with
a very high accuracy taking into account all above-mentioned
effects self-consistently, since the proton temperature is several
orders of magnitude lower than the temperature of the pickup
protons (see Fig. 3C) and even small numerical errors in the
temperature of the pickup protons can result in negative
temperatures of the solar wind protons.

Figure 5 presents the number densities and temperatures of H-atom
populations created inside (Fig.5A and 5B) and outside the
heliopause (Fig.5C and 5D). The sum of the number densities of
populations 1.0, 1.1, 1.2 is noted as 'tot1', the sum of the
number densities of populations 2.0-2.4 is noted as 'tot2'. For
comparison we present the number densities of populations 1 and 2
of B\&M model (denoted as curves 1 and 2). Curves 'tot1' and '1'
coincide, while curves 'tot2' and '2' are noticeably different.
The B\&M model overestimates the total number density of
populations 2.0-2.4. This is connected with the fact that
'temperatures' (as measures of the thermal energy) of populations
2.1-2.4 are much above the temperature of population 2 in the B\&M
model. Inside 20 AU population 1.0 dominates in the number
density, while outside the 20 AU population 2.0 becomes dominant.
Figure 6 presents differential fluxes of different populations of
H atoms at 1 AU. It is seen that different populations of H atoms
dominate  in different energy ranges. At the highest energies of
above 10 keV, population 2.2 dominates. This population consists
of atoms created in the inner heliosheath from hot pickup protons
of type 2. Population 2.1 dominates in the energy range of 0.2-6
keV . This population consists of atoms created in the inner
heliosheath from hot pickup protons of type 1. Since the both
populations are created in the inner heliosheath the measurements
of these energetic particles as planned by IBEX will provide
robust information on the properties of the inner heliosheath and,
particulary, on the behavior of pickup ions in this region.  Note
also that there is a significant difference in the ENA fluxes
predicted in the frame of one- and multi- component models.

\begin{figure}
\includegraphics[width=\linewidth]{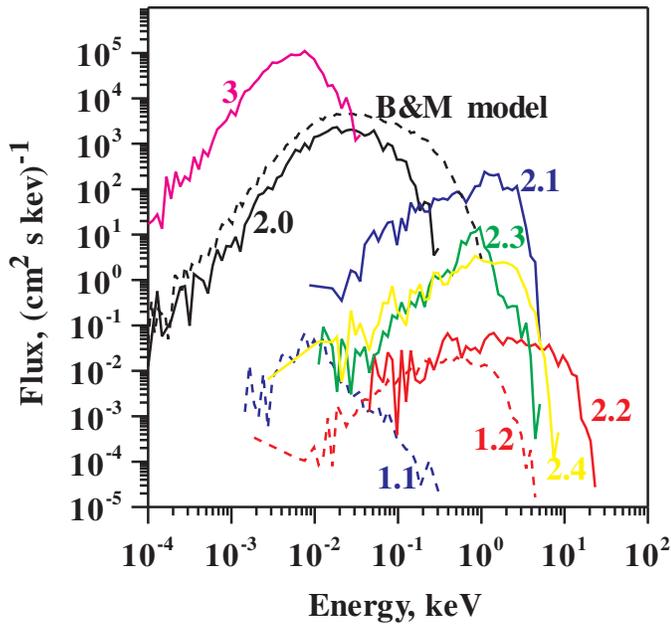}
\caption{ Fluxes of H atoms of populations 1.1,1.2, 2.0-2.4 and 3
at 1 AU in the upwind direction  as functions of energy.
\label{fig6}}
\end{figure}

Returning to Figures 5C and 5D, the filtration factor, i.e. the
amount of interstellar H atom penetrating through the interface,
and the temperature of population 3  are noticeably different in
the new multi-component model than the B\&M model. This could lead
to changes in interpretation of those observations, which require
knowledge of the interstellar H atom parameters inside the
heliosphere, say at the TS.

\section{Conclusions}
We have developed a new self-consistent kinetic-continuum model
which includes the main advantage of our previous models, i.e. a
rigorous kinetic description of the interstellar H-atom flow, and
 takes into account pickup
protons as a separate kinetic component. The new model is very
flexible and allows us to test different scenarios for the pickup
component inside, outside and at the termination shock. The model
allows to treat electrons as a separate component and to consider
different scenarios for this component. We have created a new tool
 for the interpretation of pickup ions and ENAs as
well as all diagnostics, which are connected with the interstellar
H-atom component. The new model requires a more exact description
of the physical processes involved than previous non
self-consistent models. It is shown that the heliosheath becomes
thinner and the termination shock is further from the Sun in the
new model than the B\&M model. The heliopause, however, is closer.

The main methodological advancements made in the reported model,
which was not discussed in this paper, is that we successfully
applied the Monte Carlo method with splitting of trajectories
(Malama 1991) to non-Maxwellian velocity distribution functions of
pickup protons. The splitting of trajectories allows us to
 improve the statistics of our method essentially and to calculate
differential fluxes of ENAs at 1 AU with a high level of accuracy.
We showed  that ENAs created from different types of pickup
protons dominate in different energy ranges that allows us to
determine the nature of the heliosheath plasma flow.

{\bf Acknowledgement.}  We thank our referee, Hans J. Fahr, for
valuable suggestions that improved this paper. The calculations
were performed using the supercomputer of the Russian Academy of
Sciences. This work was supported in part by INTAS Award
2001-0270, RFBR grants 04-02-16559, 04-01-00594, RFBR-CNRS (PICS)
project 05-02-22000a and Program of Basic Researches of OEMMPU
RAN. Work of V.I. was also supported by NASA grant NNG05GD69G,
RFBR-GFEN grant 03-01-39004, and International Space Science
Institute in Bern.

\end{document}